# Tailoring tricolor structure of magnetic topological insulator for robust axion insulator

M. Mogi[1]*, M. Kawamura[2], A. Tsukazaki[3], R. Yoshimi[2], K. S. Takahashi[2,4], M. Kawasaki[1,2] and Y. Tokura[1,2]*

[1]*Department of Applied Physics and Quantum Phase Electronics Center (QPEC), University of Tokyo, Bunkyo-ku, Tokyo 113-8656, Japan.*

[2]*RIKEN Center for Emergent Matter Science (CEMS), Wako, Saitama 351-0198, Japan.*

[3]*Institute for Materials Research, Tohoku University, Sendai, Miyagi 980-8577, Japan.*

[4]*PRESTO, Japan Science and Technology Agency (JST), Chiyoda-ku, Tokyo 102-0075, Japan.*

*Corresponding author. Email: mogi@cmr.t.u-tokyo.ac.jp, tokura@riken.jp

**Exploration of novel electromagnetic phenomena is a subject of great interest in topological quantum materials. One of the unprecedented effects to be experimentally verified is topological magnetoelectric (TME) effect originating from an unusual coupling of electric and magnetic fields in materials. A magnetic heterostructure of topological insulator (TI) hosts such an exotic magnetoelectric coupling and can be expected to realize the TME effect as an axion insulator. Here we designed a magnetic TI with tricolor structure where a non-magnetic layer of $(Bi, Sb)_2Te_3$ is sandwiched by a soft ferromagnetic Cr-doped $(Bi, Sb)_2Te_3$ and a hard ferromagnetic V-doped $(Bi, Sb)_2Te_3$. Accompanied by the quantum anomalous Hall (QAH) effect, we observe zero Hall conductivity plateaus, which are a hallmark of the axion insulator state, in a wide range of magnetic field between the coercive fields of Cr- and V-doped layers. The resistance of the axion insulator state reaches as high as $10^9$ Ω, leading to a gigantic magnetoresistance ratio exceeding 10,000,000% upon the transition from the QAH state. The tricolor structure of TI may not only be an ideal arena for the topologically distinct phenomena, but also provide magnetoresistive applications for advancing dissipationless topological electronics.**

**Introduction**

Three-dimensional topological insulator (TI) is a novel state of matter hosting insulating bulk and conducting surface states as protected by time-reversal symmetry (*1, 2*). When the time-reversal symmetry is broken, various exotic electromagnetic phenomena are predicted to occur (*1–3*). These include quantum anomalous Hall (QAH) effect (*3–9*), quantized magneto-optical effect (*3, 10–14*), and topological magnetoelectric (TME) effect (*3, 15–18*). The QAH effect and the quantum magneto-optical effect have already been demonstrated experimentally: The quantized Hall conductance ($e^2/h$) at zero external magnetic field was observed in the QAH effect (*5-9*), and a universal relation between the Kerr and Faraday rotation angles irrespective of material parameters was observed to converge to the fine-structure constant $\alpha = e^2/2\varepsilon_0 hc$ in the QAH/QH state (*12–14*) (Fig. 1A). These quantum phenomena are comprehensively understood as a consequence of the axion electrodynamics of TIs, named after the proposed mechanism to explain charge parity symmetry conservation of the strong interaction in particle physics (*3, 19, 20*). The TME effect is one other phenomenon derived from the axion electrodynamics, where the crossed induction of magnetization and electric polarization are expected to occur by applying external electric and magnetic fields, respectively (*3, 15–17*). Importantly, the magnetoelectric susceptibility for the crossed induction is given by the fine-structure constant $\alpha$. Despite of its impact on fundamental physics, the observation of the TME effect remains an experimental challenge.

Experimental difficulties for observing the TME effect lies mostly in finding an appropriate materials system (*21–23*). Potent examples include Weyl semimetals with strong electron correlation as exemplified theoretically in $R_2Ir_2O_7$ (*22*) and $AOs_2O_4$ (*23*) (*R*: rare earth element, *A*: alkali metal element). However, their complexity of electronic structure calls for further research. The lead candidate is the axion insulator state of a ferromagnetic three-dimensional TI, where all the surface states are insulating while maintaining the non-trivial topology of the bulk state (*3, 15–17*). The axion insulator state has recently been demonstrated by engineering a TI heterostructure thin film of (Bi, Sb)$_2$Te$_3$ (BST) with magnetic ions (Cr) doped only in the vicinity of the top and bottom surfaces of the film (*18*). As a hallmark of the axion insulator, zero Hall conductivity ($\sigma_{xy}$) plateaus (ZHP) with simultaneous vanishing of longitudinal conductivity ($\sigma_{xx}$) were reported. Anti-parallel magnetization alignment of the two magnetic layers, *i.e.*, magnetizations pointing upwards and downwards on the top and the bottom surfaces, respectively (Fig. 1B), makes all the surface states insulating to realize the axion insulator. The anti-parallel magnetization configuration arises from difference in the coercive fields ($B_c$) of the two magnetic layers. In the previous work, the difference was probably imposed by a vertical asymmetry of the heterostructure, and hence the magnetic field ($B$) for the axion insulator state was limited to a narrow range (*e.g.*, 0.12 T < $B$ < 0.15 T at 40 mK).

Towards the observation of TME effect, robust realization of the axion insulator state is strongly desirable in a wide range of external magnetic and electric fields (*24, 25*). Because the TME responses are expected to be proportional to the applied external fields, preparation of the axion insulator state in a wide range of magnetic and electric fields is advantageous to obtain large TME signals. In addition, a highly insulating state of axion insulator is preferable to prevent from screening the electric field or from relaxation of the induced polarization charge.

Here we report the realization of a robust axion insulator state by tailoring a tricolor structure of magnetic topological insulator thin film as shown in Fig. 1C. The tricolor structure was designed so that the two separated magnetic layers have different $B_c$. For the magnetic TI layers, we employed Cr (*5–7, 9*) and V (*8*) doped BST which were reported to exhibit the QAH effect. The values of $B_c$ have been reported 0.2 T and 1 T for the Cr and V doped BST, respectively. Hence if the magnetic coupling of the two layers is weak, this film structure is expected to exhibit a double step magnetization reversal with anti-parallel magnetization alignment between the two coercive fields. To ensure the magnetic decoupling of the two layers, we insert a non-magnetic BST between the Cr- and V-doped layers. The total film thickness of 9 nm was chosen to weaken the hybridization between the top and bottom surface states (*18*, *26, 27*). In this tricolor structure (Fig. 1C), electrons on the bottom surface state interact with the Cr ions and those on top surface state interact with V ions.

**Results and Disscussion**

The tricolor-structure film of Cr-doped BST (2 nm)/BST (3 nm)/V-doped BST (3 nm) (Cr-V doped BST) was grown by molecular beam epitaxy on InP(111) substrate (Fig. 1C). To align the charge neutrality points for the top and bottom surface states, Bi/Sb ratio was modulated at the Cr-doped layer and the V-doped layer (section S1 in Supplementary Materials for more details). A 1-nm-thick BST was grown as a buffer layer between the substrate and the film for the improvement of the quality, providing enhancement in the observable temperature of QAH effect (*9*). We also prepared Cr-Cr and V-V doped bicolor TI heterostructure films (insets of Fig. 1F) for comparison. Cross-sectional scanning transmission electron microscope and energy dispersive X-ray spectroscopy observations proved that Cr and V ions are distributed in the film as expected (Fig. 1D) (see also fig. S2 for more details). The films were fabricated into shapes of Hall bar and Corbino disk for transport measurements (fig. S3).

Transport measurements on the Cr-V doped BST film were conducted at 60 mK using a dilution refrigerator. Figure 1E shows the magnetic field ($B$) dependence of the Hall ($\sigma_{xy}$) and longitudinal ($\sigma_{xx}$) conductivities in a Hall bar device. Double step transitions are observed in the $B$ dependence of $\sigma_{xy}$. $\sigma_{xx}$ has double peaks at the $B$ of $\sigma_{xy}$ transitions, otherwise $\sigma_{xx}$ is almost zero. Between the two transitions in $\sigma_{xy}$, the ZHP is clearly observed in a range of 0.19 T < $B$ < 0.72 T (see also fig. S4). Figure 1F shows the transport properties of the V-V doped (3 nm) and Cr-Cr doped (2 nm) BST heterostructure at 500 mK. The $B_c$ of Cr-Cr doped and V-V doped films are 0.2 T and 0.8 T, respectively. The transition fields of the Cr-V doped film (Fig. 1E) nearly correspond to the $B_c$ of Cr-Cr doped (~ 0.2 T) and V-V doped (~ 0.8 T) heterostructures. The correspondence unambiguously shows that the ZHP originates from the anti-parallel magnetization alignment of the Cr- and V- doped layers. As shown in Fig.1F, precursors of the ZHP are seen in the $B$ dependence of $\sigma_{xy}$ in the V-V and Cr-Cr films as well. However, the anti-parallel magnetization in these films is probably induced by the unintentional difference in $B_c$ between upper and lower magnetically-doped layers, therefore the observed $B$ range for the ZHP is very narrow compared to the Cr-V film.

In Fig. 2A, we show $B$ dependence of $\sigma_{xy}$ at various temperatures from 60 mK to 20 K. A kink structure is observed around the sign changes of $\sigma_{xy}$ below 9 K, which develops into the ZHP with lowering temperature. Below 300 mK, the ZHP is well established and the quantization of $\sigma_{xy}$ to $\pm e^2/h$ is clearly seen at $B$ = 0 T. In Fig. 2B, the values of $B_c$ are plotted in the $T$-$B$ plane for the Cr-V doped and the Cr-Cr doped films. The $B_c$ is determined from peak positions of d$\sigma_{xy}$/d$B$. The smaller and the larger $B_c$ values are defined as $B_{c1}$ and $B_{c2}$, respectively. The area corresponding to the ZHP between QAH states is highlighted as a white region. Compared to the Cr-Cr doped film, the area of the ZHP is far extended in the $T$-$B$ plane as designed. $B_{c1}$ of the Cr-V doped film roughly coincides with the $B_{c1,2}$ of the Cr-Cr doped film. The area of the ZHP is extended not only along the magnetic field axis but also along the temperature axis in the Cr-V doped film. Improved stability of the anti-parallel magnetization configuration is probably responsible for the increase in the observable temperature of the ZHPs.

Next, we investigate robustness of the ZHP states against the application of electric field using two-terminal transport measurements on the Corbino disk. In Fig. 3, we show current-voltage (*I-V*) characteristics of the Cr-V doped tricolor film at 60 mK. The *I-V* curve displays an insulating behavior up to $|V| \sim 10$ mV. Application of $|V| > 10$ mV breaks down the insulating state, resulting in steep increase in *I*. We compare the *I-V* characteristics of the Cr-V doped film with those observed in the Cr-Cr doped film. The breakdown voltage of Cr-V doped film is about five times larger than that of the Cr-Cr doped film ($|V| \sim 2$ mV). Focusing on the *I-V* curves in the low voltage regime (the insets of Fig. 3), the two-terminal resistance reaches about 300 MΩ in the Cr-V doped film, which is about six times larger than that of the Cr-Cr doped film (50 MΩ). Thus, the stability of the axion insulator state against the electric field is also improved in the Cr-V doped film compared to the Cr-Cr doped film of the previous study (*18*). Although the relevant electric field direction for the TME effect is perpendicular to the film, the improved robustness against the in-plane electric field strongly suggests that the obtained axion insulator state is perhaps robust also to the perpendicular electric fields.

The large two-terminal resistance of the axion insulator state leads to a gigantic magnetoresistance ratio. Figure 4A shows *B* dependence of two-terminal resistance ($R_{2T}$) of the Hall bar device (width $W = 300$ μm and length $L = 1$ mm) at various temperatures. The $R_{2T}$ exhibits a butterfly shape when *B* is swept. The low resistance value of $R_{2T} = 28$ kΩ under the parallel magnetization configuration jumps up by about 9,000 times to 260 MΩ under the anti-parallel magnetization configuration. When the field scan is stopped at 0.38 T, the resistance value keeps increasing up to 2.8 GΩ with the elapsed time, perhaps because of the cooling down of the sample temperature which was increased during the *B* scan (fig. S5); this gives the magnetoresistance (MR) ratio [$= (R_{2T}^{AP} - R_{2T}^{P}) / R_{2T}^{P}$, where AP and P denote anti-parallel and parallel magnetization configuration, respectively] exceeding 100,000 (or 10,000,000%). The gigantic MR is demonstrated within 2 T, being distinct from the extremely large MR (XMR) reported for other topological materials, such as WTe$_2$ (*28*) and NbP (*29*), by applying high magnetic field above 50 T. Even at 0.5 K, the MR ratio remains as high as 1,800%, which is comparable to the XMR at $B = 2$ T. One other important and distinctive feature of the MR ratio in the present system is that the low resistance value $R_{2T}^{P}$ is mainly dominated by the Hall resistance. Therefore, $R_{2T}^{P}$ takes a value close to $h/e^2$ and does not strongly depend on the dimensions of the Hall bar. On the other hand, the high resistive value $R_{2T}^{AP}$ scales with the aspect ratio of the Hall bar as usual. As a result, the MR ratio associated with the QAH/axion insulator transition, which corresponds to the switching on/off of the chiral edge channels, becomes dependent on the Hall bar dimensions.

The axion insulator state with such a high resistance is maintained even when the field is swept back to $B = 0$ T. Figure 4B shows the temperature ($T$) dependence of $R_{2T}$ at $B = 0$ T under AP and P magnetization configurations ($T$ dependence of $\rho_{xx}$ and $\rho_{yx}$ measured in the Hall bar are plotted together). The resistance under AP magnetization is highly sensitive to the $T$. By increasing $T$ from 60 mK up to 400 mK, $R_{2T}^{AP}$ decreases by three-orders of magnitude. The high resistance values in Fig. 4A measured by scanning $B$ probably suffered from the temperature increase during the $B$ scan. This behavior contrasts to the $R_{2T}$ under P configuration which is mainly dominated by the Hall resistance and is only weakly dependent on $T$.

In conclusion, we have realized the robust axion insulator state by tailoring a tricolor structure of magnetic TI selectively doped with Cr and V. Observed wide ZHP width (0.19 T < $B$ < 0.72 T at 60 mK) with larger breakdown electric field are crucial improvements for the tests of TME effect. In addition, switching on/off of the chiral edge current by magnetization reversal of Cr-doped (soft ferromagnetic, $B_c$ < 0.2 T) layer provides the gigantic MR ratio exceeding 10,000,000%. Because of the scale invariant nature of the QAH state, the MR ratio in the present system depends on the sample geometry. This singularity in the MR effect of the tricolor structure of magnetic TI would carve out a new strategy for electric circuit applications of chiral edge channels, in combination with the experimental technique of arbitrary magnetic domains writing/motion (*30–32*).

**Materials and Methods**
**Thin film growth**

The TI heterostructures were grown by MBE system. The substrate was a semi-insulating (> $10^7$ Ωcm) InP(111)A and annealed at 380 °C in the chamber before growth. Bi, Sb, Te and Cr were evaporated from standard Knudsen cells, and V was evaporated from the Knudsen cell customized for a high-temperature use. For the growth of BST and Cr-doped BST, growth rate was ~0.2 nm min$^{-1}$ and growth temperature was kept at 200 °C. After the growth of intermediate BST layer, the film was annealed *in situ* at 380 °C for 30 min under exposure to Te. Subsequently, the V-doped BST layer was grown at a rate of 0.1 nm min$^{-1}$ at 160 °C. After the growth, a 3-nm-thick AlO$_x$ capping layer was immediately deposited *ex situ* by atomic layer deposition (ALD) at room temperature.

**Device fabrication**

The films were patterned to the Hall bars and Corbino disks by conventional photolithography, Ar ion-milling and chemical wet etching processes. For Ohmic contact electrodes, 3-nm-thick Ti and subsequent 27-nm-thick Au were deposited by electron-beam evaporation. The dimensions of the Hall bars are 1,000 × 300 μm which has voltage probes separated by 300 μm. The inner and outer radii of the Corbino disk are 250 μm and 400 μm, respectively.

**Transport measurement**

Electrical transport measurements of the devices were performed in a dilution refrigerator or the Quantum Design physical property measurement system (PPMS) under the control of magnetic field and temperature. Most of the measurements were performed by a standard lock-in technique with a fixed excitation current (1-10 nA) or voltage (0.1-5 mV) at low frequency (3 Hz). The *I-V* characteristic measurements were performed by a standard dc method.

**References and Notes**

1. M. Z. Hasan, C. L. Kane, Colloquium: Topological insulators. *Rev. Mod. Phys.* **82**, 3045-3067 (2010).
2. X.-L. Qi, S.-C. Zhang, Topological insulators and superconductors. *Rev. Mod. Phys.* **83**, 1057-1110 (2011).
3. X.-L. Qi, T. L. Hughes, S.-C. Zhang, Topological field theory of time-reversal invariant insulators. *Phys. Rev. B* **78**, 195424 (2008).

4. R. Yu, W. Zhang, H.-J. Zhang, S.-C. Zhang, X. Dai, Z. Fang, Quantized anomalous Hall effect in magnetic topological insulators. *Science* **329**, 61-64 (2010).
5. C.-Z. Chang, J. Zhang, X. Feng, Z. Zhang, M. Guo, K. Li, Y. Ou, P. Wei, L.-L. Wang, Z.-Q. Ji, Y. Feng, S. Ji, X. Chen. J. Jia, X. Dai, Z. Fang, S.-C. Zhang, K. He, Y. Wang, L. Lu, X.-C. Ma, Q.-K. Xue, Experimental observation of the quantum anomalous Hall effect in a magnetic topological insulator. *Science* **340**, 167-170 (2013).
6. J. G. Checkelsky, R. Yoshimi, A. Tsukazaki, K. S. Takahashi, Y. Kozuka, J. Falson, M. Kawasaki, Y. Tokura, Trajectory of the anomalous Hall effect towards the quantized state in a ferromagnetic topological insulator. *Nat. Phys.* **10**, 731-736 (2014).
7. X. Kou, S.-T. Guo, Y. Fan, L. Pan, M. Lang, Y. Jiang, Q. Shao, T. Nie, K. Murata, J. Tang, Y. Wang, L. He, T.-K. Lee, W.-L. Lee, K. L. Wang, Scale-invariant quantum anomalous Hall effect in magnetic topological insulators beyond the two-dimensional limit. *Phys. Rev. Lett.* **113**, 137201 (2014).
8. C.-Z. Chang, W. Zhao, D. Y. Kim, H. Zhang, B. A. Assaf, D. Heimann, S.-C. Zhang, C. Liu, M. H. W. Chan, J. S. Moodera, High-precision realization of robust quantum anomalous Hall state in a hard ferromagnetic topological insulator. *Nat. Mater.* **14**, 473-477 (2015).
9. M. Mogi, R. Yoshimi, A. Tsukazaki, K. Yasuda, Y. Kozuka, K. S. Takahashi, M. Kawasaki, Y. Tokura, Magnetic modulation doping in topological insulators toward higher-temperature quantum anomalous Hall effect. *Appl. Phys. Lett.* **107,** 182401 (2015).
10. J. Maciejko, X.-L. Qi, H. D. Drew, S.-C. Zhang, Topological quantization in units of the fine structure constant. *Phys. Rev. Lett.* **105**, 166803 (2009).
11. W.-K. Tse, A. H. MacDonald, Giant magneto-optical Kerr effect and universal Faraday effect in thin-film topological insulators. *Phys. Rev. Lett.* **105**, 166803 (2010).
12. K. N. Okada, Y. Takahashi, M. Mogi, R. Yoshimi, A. Tsukazaki, K. S. Takahashi, N. Ogawa, M. Kawasaki, Y. Tokura, Terahertz spectroscopy on Faraday and Kerr rotations in a quantum anomalous Hall state. *Nat. Commun.* **7**, 12245 (2016).

13. L. Wu, M. Salehi, N. Koirala, J. Moon, S. Oh, N. P. Armitage, Quantized Faraday and Kerr rotation and axion electrodynamics of a 3D topological insulator. *Science* **354**, 1124-1127 (2016).
14. V. Dziom, A. Shuvaev, A. Pimenov, G. V. Astakhov, C. Ames, K. Bendias, J. Böttcher, G. Tkachov, E. M. Hankiewicz, C. Brüne, H. Buhmann, L. W. Molenkamp, Observation of the universal magnetoelectric effect in a 3D topological insulator. *Nat. Commun.* **8**, 15197 (2017).
15. A. M. Essin, J. E. Moore, D. Vanderbilt, Magnetoelectric polarizability and axion electrodynamics in crystalline insulators. *Phys. Rev. Lett.* **102**, 146805 (2009).
16. T. Morimoto, A. Furusaki, N. Nagaosa, Topological magnetoelectric effects in thin films of topological insulators. *Phys. Rev. B* **92**, 085113 (2015).
17. J. Wang, B. Lian, X.-L. Qi, S.-C. Zhang, Quantized topological magnetoelectric effect of the zero-plateau quantum anomalous Hall state. *Phys. Rev. B* **92**, 081107(R) (2015).
18. M. Mogi, M. Kawamura, R. Yoshimi, A. Tsukazaki, Y. Kozuka, N. Shirakawa, K. S. Takahashi, M. Kawasaki, Y. Tokura, A magnetic heterostructure of topological insulators as a candidate for an axion insulator. *Nat. Mater.* **16**, 516-521 (2017).
19. R. D. Peccei, H. R. Quinn, *CP* conservation in the presence of pseudoparticles. *Phys. Rev. Lett.* **38**, 1440-1443 (1977).
20. F. Wilczek, Two applications of axion electrodynamics. *Phys. Rev. Lett.* **58**, 1799-1802 (1987).
21. R. Li, J. Wang, X.-L. Qi, S.-C. Zhang, Dynamical axion field in topological magnetic insulators. *Nat. Phys.* **7**, 284-288 (2010).
22. X. Wan, A. M. Turner, A. Vishwanath, S. Y. Savrasov, Topological semimetal and Fermi-arc surface states in the electronic structure of pyrochlore iridates. *Phys. Rev. B* **83**, 205101 (2011).
23. X. Wan, A. Vishwanath, S. Y. Savrasov, Computational design of axion insulators based on 5*d* spinel compounds. *Phys. Rev. Lett.* **108**, 144601 (2012).
24. M. Fiebig, Revival of the magnetoelectric effect. *J. Phys. D: Appl. Phys.* **38**, R123 (2005).
25. J.-P. Rivera, A short review of the magnetoelectric effect and related experimental techniques on single phase (multi-) ferroics. *Eur. Phys. J. B.* **71**, 299 (2009).

26. C.-X. Liu, H. J. Zhang, B. Yan, X.-L. Qi, T. Frauenheim, X. Dai, Z. Fang, S.-C. Zhang, Oscillatory crossover from two dimensional to three dimensional topological insulators. *Phys. Rev. B* **81**, 041307(R) (2010).

27. Y. Zhang, K. He, C.-Z. Chang, C.-L. Song, L.-L. Wang, X. Chen, J.-F. Jia, Z. Fang, X. Dai, W.-Y. Shan, S.-Q. Shen, Q. Niu, X.-L. Qi, S.-C. Zhang, X.-C. Ma, Q.-K. Xue, Crossover of the three-dimensional topological insulator $Bi_2Se_3$ to the two-dimensional limit. *Nat. Phys.* **6**, 584-588 (2010).

28. M. N. Ali, J. Xiong, S. Flynn, J. Tao, Q. D. Gibson, L. M. Schoop, T. Liang, N. Haldolaarachchige, M. Hirschberger, N. P. Ong, R. J. Cava, Large, non-saturating magnetoresistance in $WTe_2$. *Nature* **514**, 205-208 (2014).

29. C. Shekhar, A. K. Nayak, Y. Sun, M. Schmidt, M. Nicklas, I. Leermakers, U. Zeitler, Y. Skourski, J. Wosnitza, Z. Liu, Y. Chen, W. Schnelle, H. Borrmann, Y. Grin, C. Felser, B. Yan, Extremely large magnetoresistance and ultrahigh mobility in the topological Weyl semimetal candidate NbP. *Nat. Phys.* **11**, 645-649 (2015).

30. A. L. Yeats, P. J. Mintun, Y. Pan, A. Richardella, B. B. Buckley, N. Samarth, D. D. Awschalom. Local optical control of ferromagnetism and chemical potential in a topological insulator. Preprint at https://arxiv.org/abs/1704.00831 (2017).

31. Y. Fan, P. Upadhyaya, X. Kou, M. Lang, S. Takei, Z. Wang, J. Tang, L. He, L.-T. Chang, M. Montazeri, G. Yu, W. Jiang, T. Nie, R. N. Schwartz, Y. Tserkovnyak, K. L. Wang, Magnetization switching through giant spin-orbit torque in a magnetically doped topological insulator heterostructure. *Nat. Mater.* **13**, 699-704 (2014).

32. K. Yasuda, A. Tsukazaki, R. Yoshimi, K. Kondou, K. S. Takahashi, Y. Otani, M. Kawasaki, Y. Tokura, Current-nonlinear Hall effect and spin-orbit torque magnetization switching in a magnetic topological insulator. Preprint at http://arxiv.org/abs/1612.06862 (2016).

**Acknowledgments**: We thank S. Shimizu for experimental support.

**Funding:** This research was supported by the Japan Society for the Promotion of Science through the Funding Program for World-Leading Innovative R & D on Science and Technology (FIRST Program) on "Quantum Science on Strong Correlation" initiated by the Council for Science and Technology Policy, JSPS/MEXT Grant-in-Aid for Scientific Research (No. 24224009, 24226002, 15H05867, 15H05853), and JST CREST (No. JPMJCR16F1). M.M. is supported by JSPS through a research fellowship for young scientists (No. 17J03179).

**Author contributions:** Y.T. and M.M. conceived and designed the research. M.M. grew materials and performed measurements and the analysis with the help of M.Kawamura, R.Y. and K.S.T. M.M., M.Kawamura, A.T., M.Kawasaki and Y.T. interpreted the results and wrote the manuscript with contributions from all authors.

**Competing interests:** The authors declare no competing financial interests.

**Data and materials availability:** All data needed to evaluate the conclusions in the paper are present in the paper and/or the Supplementary Materials. Additional data are available from authors upon request.

**Figures and Tables**

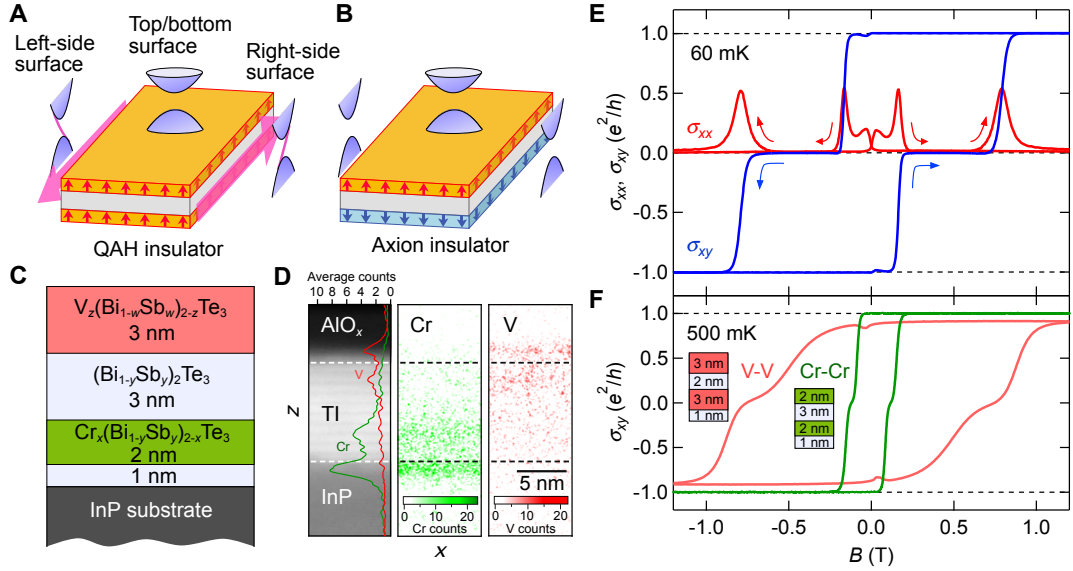

**Fig. 1. Tailoring tricolor structure of magnetic TI and robust zero Hall conductivity plateau.** (**A** and **B**) Schematic depictions of QAH and axion insulator states. Magnetization opens a gap in top and bottom surface states. When the magnetizations (arrows) are parallel (A), 1D chiral edge channels are formed on the side surface. When the magnetizations are anti-parallel (B), an energy gap opens also on the side surface. (**C**) Schematic structure of MBE-grown TI (BST) film doped with Cr and V used in the present study. The compositions: $(x, y, z, w) = (0.24, 0.74, 0.15, 0.80)$. (**D**) Cross-sectional annular dark field scanning transmission electron microscopy (ADF-STEM) image (left) and net-count mappings of Cr (center) and V (right) obtained by an energy dispersive X-ray spectroscopy (EDX) of the Cr-V doped tricolor BST film, which are taken in the same area of the ADF-STEM image. $x$ and $z$ indicate in-plane and out-of-plane directions of the film. Black bar indicates 5 nm in length. On the ADF-STEM image, averages along $x$ of net counts of Cr and V are indicated by green and red, respectively. (**E**) Magnetic field ($B$) dependence of Hall conductivity ($\sigma_{xy}$) and longitudinal conductivity ($\sigma_{xx}$) of the Cr-V doped BST film at 60 mK. (**F**) Magnetic field ($B$) dependence of $\sigma_{xy}$ at 500 mK in the Cr-Cr doped [$Cr_{0.24}(Bi_{0.26}Sb_{0.74})_{1.76}Te_3$] (shown in green) and the V-V doped [$V_{0.15}(Bi_{0.20}Sb_{0.80})_{1.85}Te_3$] (shown in red) bicolor TI heterostructures. Their growth profiles are depicted as insets with the respective thicknesses of layers.

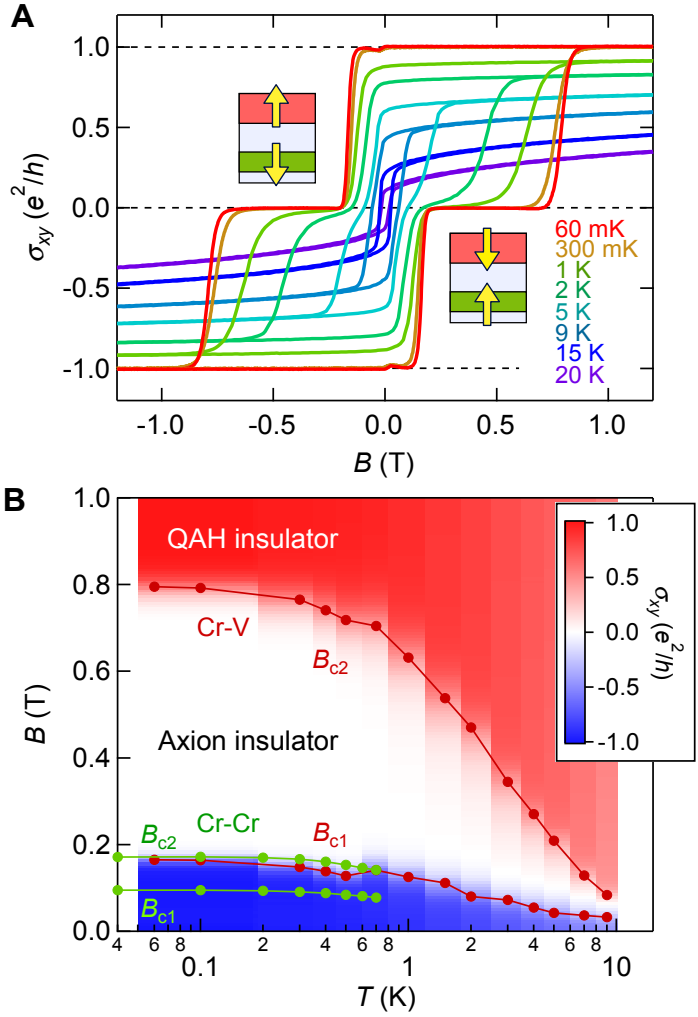

**Fig. 2. Magnetic field and temperature dependence of zero $\sigma_{xy}$ plateau.** (**A**) Magnetic field dependence of $\sigma_{xy}$ at representative temperatures in the Cr-V doped tricolor TI film. (**B**) Color mapping of $\sigma_{xy}$ for the Cr-V doped TI tricolor film and plots of peaks for $d\sigma_{xy}/dB$, which correspond to coercive fields (defined as $B_{c1}$ for smaller and $B_{c2}$ for larger) of the Cr-V doped tricolor film (shown in red) and the Cr-Cr doped bicolor film (shown in green) in $T$-$B$ plane.

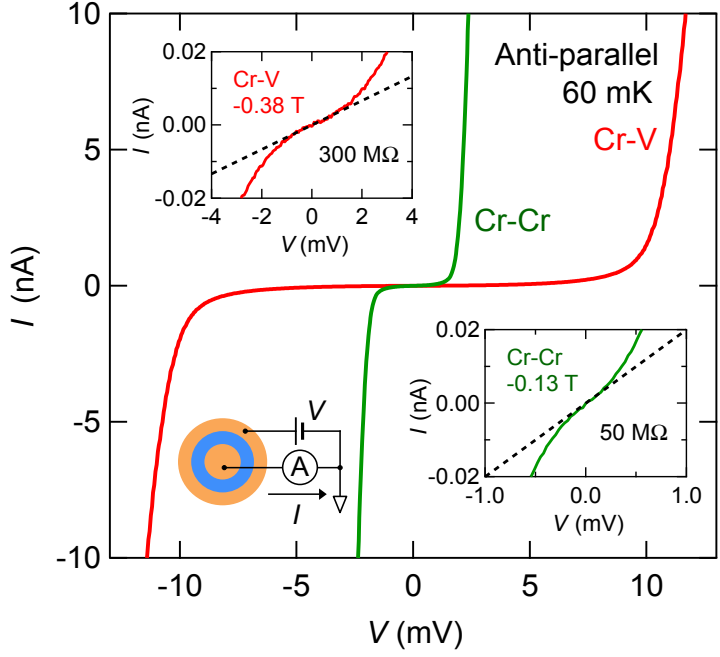

**Fig. 3. *I-V* characteristics in tricolor and bicolor TI heterostructures.** *I-V* curves under the anti-parallel magnetization configuration of the Cr-V doped tricolor BST (red) and the Cr-Cr doped bicolor BST (green) heterostructures, taken in a Corbino disk device at *T*~50 mK. For stabilization of the anti-parallel magnetization configuration, the external magnetic field is fixed at the minimum of conductivity (for the Cr-V doped BST, $B = -0.38$ T and for the Cr-Cr BST, $B = -0.13$ T). Upper (lower) inset: expanded scale of the *I-V* curve of the Cr-V (Cr-Cr) doped BST heterostructure.

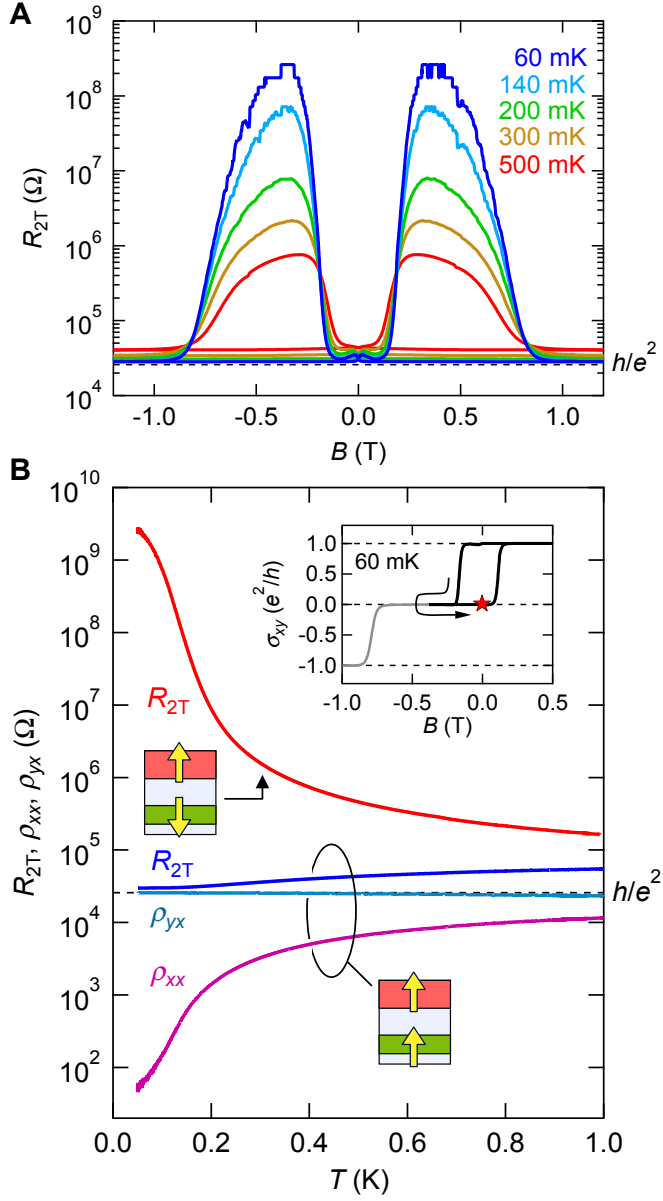

**Fig. 4. Magnetic field and temperature dependence of two-terminal resistance in Cr-V doped tricolor TI.** (**A**) Magnetic field ($B$) dependence of two-terminal resistance ($R_{2T}$) between current terminals in the Hall bar of the Cr-V doped BST film measured at various temperatures ($T$ = 60, 140, 200, 300, 500 mK). (**B**) Temperature ($T$) dependence of $R_{2T}$ in the parallel/anti-parallel magnetization configuration under zero magnetic field. $\rho_{yx}$ and $\rho_{xx}$ measured by four-terminal probe are also shown. Inset: minor-loop-hysteresis of $\sigma_{xy}$ at $T$ = 60 mK, reversing $B$ at –0.38 T.